\documentstyle[aps,twocolumn,epsf]{revtex}

\begin{document}
\title{
Bond-versus-site doping models for off-chain-doped
Haldane-gap system Y$_2$ Ba Ni O$_5$ 
}
\author{Jizhong Lou}
\address{
Department of Physics, Peking University, 
Beijing 100871, P.R. China
}
\address{
and Institute of Theoretical Physics, P. O. Box 2735, Beijing
100080, P.R. China
}
\author{Shaojin Qin and Zhaobin Su}
\address{
Institute of Theoretical Physics, P. O. Box 2735, Beijing
100080, P.R. China
}
\author{Lu Yu}
\address{
International Center for Theoretical Physics, P. O. Box 586,
34100 Trieste, Italy
}
\address{
and Institute of Theoretical Physics, P. O. Box 2735, Beijing
100080, P.R. China
}
\date{ \today }

\maketitle

\begin{abstract}
Using the density matrix renormalization-group technique,  
we calculate the impurity energy levels for  two different 
effective
models of off-chain doping for quasi-one-dimensional Heisenberg chain
compound Y$_2$ Ba Ni O$_5$: ferromagnetic bond doping and 
antiferromagnetic site spin-$1/2$ doping.  
Thresholds of the impurity strength for the
appearance of localized states are 
found for both models. 
However, the ground-state and low-energy excitations for weak impurity
strength are different for these two models and the difference can be
detected by experiments.
\end{abstract}

\medskip




In the last fifteen years many theoretical 
and experimental efforts have 
been devoted to
testing Haldane's original conjecture${}^{\cite{Haldane}}$ that
the excitation spectrum 
for integer spin Heisenberg antiferromagnetic (AF) chains has a finite 
energy gap above the ground state while it is gapless for 
half-integer cases. 
The experiments on NENP and Y$_2$ Ba Ni O$_5$ have shown explicitly 
the gap's existence.  The  gap value can also be estimated 
by numerical studies, e.g., for isotropic spin-1 AF Heisenberg 
chain, a gap $\Delta \approx$ 0.41$J$ (where $J$ is the exchange
integral)  has  been obtained by 
density matrix renormalization-group ${}^{\cite{White},\cite{WN}}$ (DMRG)
technique as well as 
by exact diagonalization with proper extrapolation${}^{\cite{SHANK}}$.
However, the knowledge of  these
spin chains is still far from complete, especially regarding the 
impurity effects.  Recently, the interest in the
impurity problem has been revived due to the doping experiment 
performed on the linear chain compound
Y$_2$ Ba Ni O$_5$ ${}^{\cite{DiTusa},\cite{KOJIMA}}$. 
Several theoretical studies have been carried out to study the
doping effect${}^{\cite{SA1,WW,WXQ}}$. However, the off-chain
doping (when Y$^{3+}$ is replaced by Ca$^{2+}$ or Mg$^{2+}$)
has not been fully studied.  This is the subject of the present
paper.

Y$_2$ Ba Ni O$_5$ contains one-dimensional chain structures
composed of oxygen octahedra with Ni$^{2+}$ ions at the center.
The inelastic-neutron-scattering (INS) experiments on 
single crystals of this charge transfer insulator show the chains
are highly one dimensional with very weak interchain coupling 
$J_2$ ($J_2 / J \leq$ 5 $\times$ 10$^{-4}$),${}^{\cite{Xu}}$  
where $J$  is the intrachain coupling,
and the Haldane gap
at $\tilde{q} = \pi$ is 
almost isotropic, so it is an ideal substance to confront
experiment.  Furthermore, the doping of this 
compound can be in-chain as well as off-chain.  The nonmagnetic 
Zn$^{2+}$ ions substituting Ni$^{2+}$ will effectively severe 
the chain and produce free s=1/2 spins at the two ends, 
which can be explained by the valence bond 
solid (VBS) model.${}^{\cite{AKLT}}$ 
Though the specific-heat measurements displaying 
Schottky anomaly${}^{\cite{RAMIREZ}}$ seemed to contradict 
this picture, a recent DMRG 
numerical study shows this conflict can be 
resolved by taking into account the 
chain length distribution.${}^{\cite{HALLBERG}}$
On the other hand,
the off-chain substitution of Y$^{3+}$ by Ca$^{2+}$ or Mg$^{2+}$ 
will induce holes along the chain.
The INS experiments show the holes 
are localized mainly on the oxygen 
sites within the chains,${}^{\cite{DiTusa}}$  
between the nearest two Ni$^{2+}$ ions.  
Since the 
coupling of the two nearby Ni$^{2+}$ ions is mediated by this 
oxygen, the doping 
affect can be modeled in two different ways:
one is that the coupling of the 
two Ni$^{2+}$ is modified (bond doping), 
the other is that an s=1/2 spin will be localized at the oxygen site, 
and the localized spin will be 
coupled to the two Ni$^{2+}$ (s=1/2 
site doping).  The Hamiltonian of these two effective doping models 
can be written as

\begin{equation}
H_{bd} = J \sum_{i=1}^{N-1} {\bf {S}}_{i} \cdot {\bf {S}}_{i+1} 
        + J^{\prime} {\bf {S}_{1}} \cdot {\bf {S}}_{N} 
\label{HBD}
\end{equation}
and
\begin{equation}  
H_{sd} = J \sum_{i=1}^{N-1} {\bf {S}}_{i} \cdot {\bf {S}}_{i+1} 
	+ J^{\prime \prime} ({\bf {S}}_{1} \cdot {\bf {\sigma}}_{0} 
			+ {\bf {\sigma}}_{0} \cdot {\bf {S}}_{N}),
\label{HSD}
\end{equation}
respectively.  Here the periodic boundary condition (PBC) is
adopted, $N$ is the chain length,  
$\bf{S}$ denotes the s=1 spin; and in Eq.(\ref{HSD}) ,
$\bf{\sigma}_{0}$ 
represents the s=1/2 impurity , and here we have
assumed that the coupling of the two Ni$^{2+}$ ions around the
impurity are destroyed completely.  

From the site-doping point of view, the induced s=1/2 
spin couples antiferromagnetically with Ni$^{2+}$ 
on both sides of it.
Pictorially, the 
spin direction of the two Ni spins around 
the impurity site will 
be antiparallel to the direction of the localized s=1/2 spin, 
that is, the two s=1 Ni spins tends to be in the same direction. 
So the effective coupling of the two site will be ferromagnetic 
instead of antiferromagnetic.  
Generically, in the Hamiltonian (\ref{HBD}), the 
coupling $J^{\prime}$ will be negative, while for the site 
doping model in Hamiltonian (\ref{HSD}), 
$J^{\prime \prime}$ will 
be positive. 

Several numerical studies have been carried out for the two 
Hamiltonians (\ref{HBD}) and 
(\ref{HSD}),$^{\cite{SA1,WW,WXQ,L1,L2}}$ 
but nearly all 
the calculations were done under 
the assumption that the impurity 
bond remains antiferromagnetic.  
In this communication, 
we use the DMRG algorithm to calculate the energy spectrum for the two
doping models (\ref{HBD}) and (\ref{HSD}), 
especially for ferromagnetic bond 
doping model.  
If $J^{\prime}$ (negative) is weak enough in
Hamiltonian (\ref{HBD}), an in-gap bound state is localized around 
the impurity site.  There is a threshold value $J^{\prime}_c$,
above which there is no in-gap level or corresponding localized
state.  We also calculate the spectrum for the impurity state for
antiferromagnetic $J^{\prime \prime}$ in 
Hamiltonian (\ref{HSD}).
Combined with previous studies of other authors$^{\cite{L1,L2}}$,
the two Hamiltonians with $J^{\prime}$ and $J^{\prime \prime}$ 
AFM as well as 
FM coupling give a
complete description for the spin-1 Heisenberg chain
doping models.  
Experimental data may distinguish which of the two 
models is more appropriate in 
describing the off-chain doped Y$_2$ Ba Ni O$_5$ compound.

The DMRG scheme we used is similar to the one used in 
Ref.\cite{WXQ}, we keep 250 states up to chain length 50 
throughout all the calculation, and the largest truncation error in our
calculation is of the order of $10^{-6}$.  At last the Shanks 
transformation$^{\cite{SHANK}}$ is used to obtain the final result. 
This transformation has been proved to be valid for spin-1 
Heisenberg chain by several 
authors${}^{\cite{SHANK,WW,WXQ}}$, and we can trust
the accuracy of this scaling approach.
Before presenting the numerical results, 
a physically intuitive
analysis of the two doping models will be useful.

First, for Hamiltonian (\ref{HBD}) of the bond doping cases, 
results in two limiting cases ($J^\prime\to 0^-$ and  
$J^\prime\to -\infty$) can be obtained within the framework of the 
VBS picture.  
When $J^{\prime}=0$, the system becomes an open Heisenberg chain, 
and the ground state will be fourfold degenerate, where the two 
edge $s=\frac{1}{2}$ spins couple into a singlet and a triplet. 
The lowest excited state with $S^{z}_{tot}=2$ is separated from
the ground state by the Haldane gap, and the continum 
spectrum starts there.  When the impurity bond intensity increases 
gradually ($\mid J^{\prime}\mid<<J, J^{\prime}< 0$ in perturbation
regime), the two edge spins will tend to form a triplet, 
while the singlet 
state will enter the Haldane gap and become an in-gap state 
between the triplet ground state and the continuum spectrum.  
The perturbative result obtained by S\o rensen and Affleck 
${}^{\cite{SA1}}$ (SA) is valid also for 
the negative $J^{\prime}$ case. 
On the other hand, for the strong impurity doping strength, 
$J^{\prime} \rightarrow -\infty$, the two $s=1$ spins at
the ends of
 the impurity bond cannot be reduced to $s=\frac{1}{2}$ spins 
anymore.  We must consider the reduced four sites model 
near the impurity bond and write 
down the effective Hamiltonian as
\begin{equation}
H=J_{eff} {\bf {\sigma}}_{1} \cdot {\bf {S}}_{1} 
	+ J^{\prime} {\bf {S}}_{1} \cdot {\bf {S}}_{2} 
	+ J_{eff} {\bf {S}}_{2} \cdot {\bf {\sigma}}_{2} \;\;,
\label{RBD}
\end{equation}
where $\bf{S}_{1}$ and $\bf{S}_{2}$ are $s=1$ spins, and 
$\bf{\sigma}_{1}$ and $\bf{\sigma}_{2}$ are $s=\frac{1}{2}$ spins 
due to the VBS picture (the remaining part of the 
periodic chain can be
viewed as a length L - 2 open chain).
  
\begin{figure}[hbt]
\epsfxsize=\columnwidth
\epsfbox{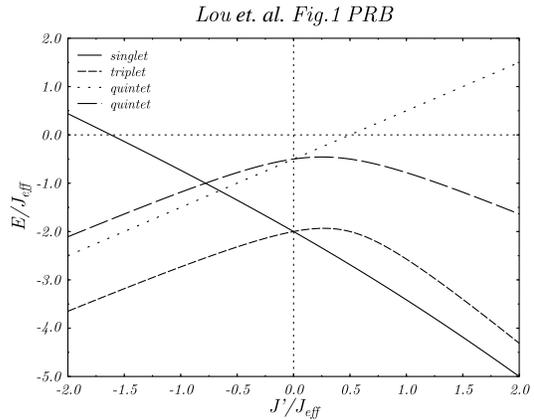}
\caption{
The low-energy spectrum of the 
reduced bond doping Hamiltonian (\ref{RBD}). 
 }
\label{fig1}
\end{figure}

In fact, the model Hamiltonian (\ref{RBD}) can account  
for the essential features of  the 
Hamiltonian (\ref{HBD}) for all values 
of $J^{\prime}$, from positive to 
negative beyond the assumption that $J_{eff}$ is positive.
Its  spectrum  contains two singlets, four triplets, there quintets 
and one septet.  Out of the ten states, only a few are 
important to our discussion: the ground state for $J^{\prime} >0$ 
and that for $J^{\prime} <0 $,  the quintets,  which are the low 
excitation states for some values of $J^{\prime} < 0$.  The energy 
of these  states versus the impurity coupling $J^{\prime}$ 
(in units of $J_{eff}$) is shown in Fig. \ref{fig1}. 

For the $J^{\prime}<0$ case on the left part of Fig. \ref{fig1}, we 
see that the ground state is a triplet for all $J^{\prime}$. 
When $J^{\prime} \rightarrow -\infty$, the first excited state 
is a quintet instead of a singlet.  This is an interesting 
phenomenon for the ferromagnetic impurity doping case, 
that is, there will be a possible energy crossover between the singlet and 
quintet for larger $|J^{\prime}|$. 
Since the energy difference of this quintet to the ground state 
is nonzero, we can expect either one bound state for small 
$J_{eff}$  (more if $J_{eff}$ is small enough) or no bound states 
for larger $J_{eff}$ within the bulk gap 
$\Delta=0.41048(2)$.  We will show later that for large $|J^{\prime}|$
there is no such crossover from our 
DMRG calculation, and there is no in-gap 
states for long chains. Therefore, $J_{eff}$ is large enough.
Since we know very little about $J_{eff}$ at the present stage, it is 
impossible to conclude which case will occur without more detailed 
computation.

For $J^{\prime}>0$, the AFM doping case on the right part of Fig. \ref{fig1}, 
only two states are important: the singlet ground state and the first 
triplet excitation state.  For $J^{\prime}=0$ and $J^{\prime} 
\rightarrow \infty$, the energy difference is zero, so the gap state 
can be expected for both weak and strong impurity bond strength.  
Detailed numerical results for this case can be found in 
previous studies.${}^{\cite{SA1,WW,WXQ}}$

\begin{figure}[b]
\epsfxsize=\columnwidth
\epsfbox{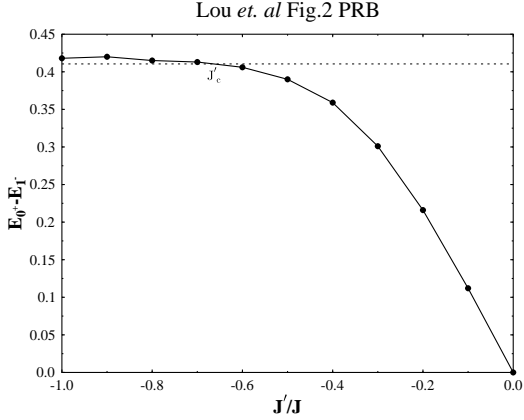}
\caption{
The energy spectrum of the bond 
doping Hamiltonian (\ref{HBD}) when the
impurity bond is ferromagnetic. The dotted line 
indicates the Haldane gap of a 
pure spin-1 Heisenberg chain.
}
\label{fig2}
\end{figure}

Now we present DMRG results for Hamiltonian (\ref{HBD}) with FM bond
doping. 
The dependence of the relative energy of the lowest excited state
with respect
to the ground state on the impurity bond is shown in Fig. \ref{fig2}.  From 
the figure, we see clearly that a threshold $J^\prime_c$ exists 
above which the in-gap state disappears.  For weaker impurity bond, 
the impurity level increases linearly with the bond strength, 
which agrees with 
SA's perturbative argument$^{\cite{SA1}}$ 
for the weak coupling limit. 
The relation between the splitting of the ground state and 
the singlet excitations is $\Delta E=-\alpha^{2} J^{\prime}$, 
where $\Delta E$= 0.112$J$ for $J^{\prime}$=-0.1$J$, 
which is also 
in excellent agreement with the theoretical work
for the AFM case${}^{\cite{SA1}}$.  When the impurity 
bond strength becomes bigger, the energy of the in-gap 
state deviates from linear and enters 
the continuum spectrum at last 
at a critical point for bond strength near $J^{\prime} = -0.7J$. 
The simple four-site model based on the VBS picture discussed above
shows a crossover between the singlet state (in-gap state 
for some values of $J^{\prime}$ ) and the quintet 
state, which remains at
the bottom of the continuum spectrum.  we did not find such a 
feature in 
the calculation, so there is no energy crossover taking place,
this is due to the interaction between the singlet state and
the continuum spectrum, which keeps the singlet state as the
lowest-energy excitation state for longer chain length.

\begin{figure}[hbt]
\epsfxsize=\columnwidth
\epsfbox{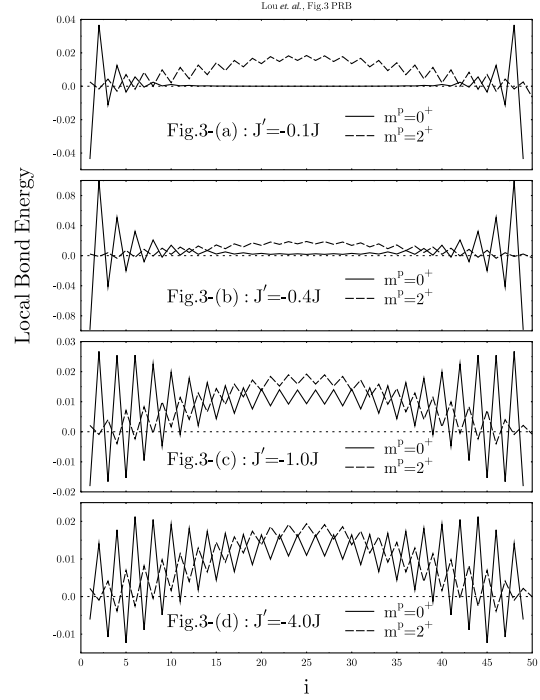}
\caption{
The local bond energy 
$\langle {\bf S}_i \cdot {\bf S}_{i+1} \rangle_{m^p}
-\langle {\bf S}_i \cdot {\bf S}_{i+1} \rangle_{1^-}$ 
for different ferromagnetic 
doping bond $J^{\prime}$, 
where $m^p$ equals $0^+$ and $2^+$.
}
\label{fig3}
\end{figure}

The local bond energy 
$\langle {\bf S}_i \cdot {\bf S}_{i+1} \rangle_{m^p}-
\langle {\bf S}_i \cdot {\bf S}_{i+1} \rangle_{1^-}$ 
for different impurity bond strengths is presented in Fig. \ref{fig3} 
(here $m$ represent the total spin and $p$ is the parity of the state).
The $m^p=2^+$ state corresponds to the 
one-magnon excitation${}^{\cite{SA2,SA3}}$. It change very little
when the impurity bond strength changes from very weak (-0.1$J$)
up to very strong (4.0$J$)(this is different from the 
positive $J^{\prime}$ case, because for $J^{\prime}$=1.0$J$, this is
a two-magnon states${}^{\cite{SA2,SA3}}$). While the 
$m^p=0^+$ 
state is an edge excitation localized around the impurity bond for 
weak $J^{\prime}$.  When $J^{\prime}$ becomes bigger, this 
excitation is delocalized and shows a single-magnon-like 
behavior in the strong coupling region. This is also consistent
with the absence of the in-gap impurity states in this region.

\begin{figure}[hbt]
\epsfxsize=\columnwidth
\epsfbox{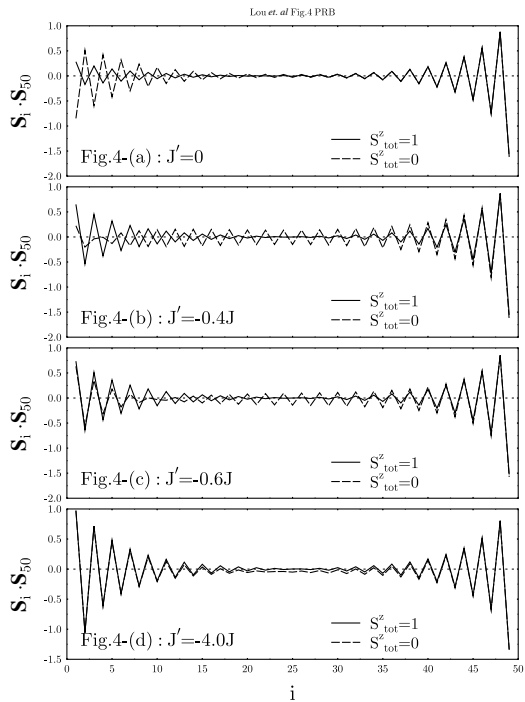}
\caption{
The pair correlations 
$\langle {\bf S}_i \cdot {\bf S}_{50} \rangle$
 of the ferromagnetic bond doping for the 
 ground state $1^-$ and the first excitation
 state $0^+$.
 }
\label{fig4}
\end{figure}

We also calculate the ground state and the first excited state 
(the subgap state) pair correlations of the site nearest to the 
impurity bond for different $J^{\prime}$ , that is, 
$\langle {\bf S}_i \cdot {\bf S}_{50} \rangle$, which is presented in 
Fig. \ref{fig4}.  For $J^{\prime}=0$, the open chain case, we see that for
$i > 25 $, the pair correlations for the two states are in phase,
but for $i < 25 $, they are out of phase. This phase difference 
which keeps the orthogonality of the two states
is due to the different coupling of the two chain-end 1/2 spins. 
In contrast, for 
strong $J^{\prime}$, the correlations  are almost identical 
except those $i$ in the middle of the chain where the one-magnon
amplitude is the biggest.${}^{\cite{SA2,SA3}}$ The correlations
for the two states are both independent of $J^{\prime}$ for $i > 25$, 
while for $S^z_{tot}$=1, the ground state, the correlations for
$i < 25$ will become larger, and for $S^z_{tot}=0$, the first excited
state, the correlation will change its phase when 
increasing $J^{\prime}$, the orthogonality of the two states will mainly
come from those sites in the middle of the chain, this also means that the
edge excitation (near the impurity bond) when $J^{\prime}$ is weaker will 
become the bulk excitation (far away from the impurity bond)
when $J^{\prime}$ is stronger.

Secondly, we analyze the spectrum of Hamiltonian (\ref{HSD}). 
The impurity coupling $J^{\prime \prime}$ is positive.  
Let us reduce 
the Hamiltonian (\ref{HSD}) to a three sites effective Hamiltonian 
around the impurity $s=1/2$:
\begin{equation}  
H =  J_{eff} ({\bf {\sigma}}_{1} \cdot {\bf {\sigma}}_{0} 
		+ {\bf {\sigma}}_{0} \cdot {\bf {\sigma}}_{N}).
\label{RSD}
\end{equation}
Here $\bf{\sigma}_{1}$ and $\bf{\sigma}_{N}$ are the
effective $s=1/2$ edge spins from the bulk of the chain, and 
$\bf{\sigma}_{0}$ is the impurity spin.  As we analyzed above,  
 here we have $J_{eff}>0$.  In the perturbation regime 
($J_{eff}\sim 0$), we know that Hamiltonian (\ref{RSD}) has a doublet 
ground state, a doublet first excited state, and a quartet highest
excited state.${}^{\cite{SA1}}$  
For the original Hamiltonian (\ref{HSD}), as 
$J_{eff}$ increases, we could see that the doublet ground state
will not change, while the two in-gap 
states given by the reduced
Hamiltonian increase linearly with $J_{eff}$.  By DMRG calculation,
we find $J^{\prime \prime}$ in Eq. (\ref{HSD}) 
has two critical impurity coupling 
$J^{\prime \prime}_{1c}$ and $J^{\prime \prime}_{2c}$,
which  correspond to the 
points where the doublet and the quartet
enter the continuum spectrum, respectively.

\begin{figure}[hbt]
\epsfxsize=\columnwidth
\epsfbox{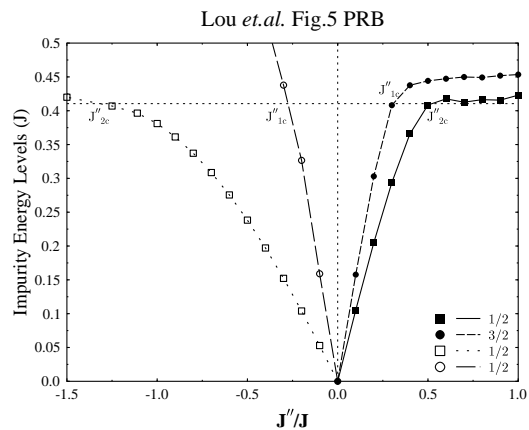}
\caption{
The energy spectrum of the site 
doping Hamiltonian (\ref{HSD}). The dotted 
line indicates the Haldane gap of 
the pure spin-1 Heisenberg chain.
}
\label{fig5}
\end{figure}

For the site doping Hamiltonian (\ref{HSD}), we consider both the 
ferromagnetic coupling ($J^{\prime \prime} < 0$) and the 
antiferromagnetic coupling case ($J^{\prime \prime} > 0$). 
The ground state 
is a quartet for the former while a doublet for the latter. 
Two in-gap 
states are found in the weak coupling regime in each of the two cases, 
and its energy spectrum  is shown in Fig. \ref{fig5}. 
The thresholds where the
in-gap state enters the continuum 
are $J^{\prime \prime}_{1c}=-0.25 J$, 
$J^{\prime \prime}_{2c}= -1.4 J$ and 
$J^{\prime \prime}_{1c}=0.3 J$, 
$J^{\prime \prime}_{2c}=0.5 J$ for the 
ferromagnetic and antiferromagnetic impurity coupling, respectively. 
These results are consistent with the previous 
calculation.${}^{\cite{SA1}}$

In summary, we have considered two kinds
of effective models for off-chain
doping in a one-dimensional Heisenberg chain using the DMRG 
algorithm.  Thresholds of the impurity bond strength are found for 
both ferromagnetic bond doping model and antiferromagnetic $s=1/2$ 
site doping model.  The ground states of the two kinds of effective 
models are of total spin $1$ and total spin $1/2$, respectively.  
The in-gap states  that appear 
for the weak impurity bond are also different for the two models.
A singlet impurity state for ferromagnetic bond doping and 
a doublet along with a quartet for AF site $s=1/2$ doping.  
We believe
these properties of the spectrum for different effective doping
models can be 
distinguished by experiments.
The doped holes in the off-chain doping case can contribute to
the transport properties of the Haldane chain systems. This
issue was discussed by other authors${}^{\cite{L2}}$ and is not
addressed here.

This work was supported by the NSFC and the calculations 
were supported by the LSEC 
and CCAST.  One of the authors (J.L.) would like to thank Dr. T. Xiang 
and Dr. X. Wang
for useful discussions.

\end{document}